\documentclass[twocolumn,showpacs,preprintnumbers,amsmath,amssymb,superscriptaddress]{revtex4}
\usepackage[cp1251]{inputenc}
\usepackage[ukrainian]{babel}
\usepackage{bm}
\usepackage{epsfig,amssymb,amsmath,bm}
\usepackage[dvips]{color}

\begin{document}

\title{The spin group of fermions}
\author{K. S. Karplyuk}
\email{karpks@hotmail.com}
 \affiliation{Department of Radiophysics, Taras Shevchenko University, Academic
Glushkov prospect 2, building 5, Kyiv 03122, Ukraine, karpks@hotmail.com}
\author{O. O. Zhmudskyy}\email{ozhmudsky@physics.ucf.edu}
 \affiliation{Department of Physics, University of Central Florida, 4000 Central Florida Blvd. Orlando, FL, 32816, ozhmudsky@Knights.ucf.edu}
\begin{abstract}
It was shown that in the small Wigner group there is a one-parameter subgroup of the Lorentz transformations, which leave unchanged not only the momentum of the fermion with spin $\hbar/2$, but also its spin characteristics. This is the group of symmetry transformations of the moving fermion. These transformations consist of rotations around the projection axis of the spin and boosts coordinated with them. In other words, there are infinitely many frames of reference in which the fermion has the same momentum and the same spin projection onto the selected axis, although these frames move relative to each other and have been rotated relative to each other.
\end{abstract}

\maketitle

\section{Introduction}
It is known that a fixed fermion with spin projection $\frac{\hbar}{2}$ onto the direction of the three-dimensional unit pseudovector $\bm{\mathfrak{s}}$ ($\mathfrak{s}^2_x+\mathfrak{s}^2_y+\mathfrak{s}^2_z=1$) is described by the bispinor $\psi^{(0)}$, which is the solution to the Dirac equation for the
 fermion at rest
\begin{gather}
p_0\gamma^0\psi^{(0)}=mc\psi^{(0)}
\end{gather}
and is a eigen bispinor of the operator $\frac{\hbar}{2}\frac{i\mathfrak{s}_{kl}\sigma^{kl}}{2}$:
\begin{gather}
\frac{\hbar}{2}\frac{i\mathfrak{s}_{kl}\sigma^{kl}}{2}\psi^{(0)}=\frac{\hbar}{2}\psi^{(0)}.
\end{gather}
Here $\sigma^{kl}=\frac{1}{2}(\gamma^k\gamma^l-\gamma^l\gamma^k)$, $\mathfrak{s}_{kl}$ are coefficients antisymmetric in $k$  and  $l$, $\mathfrak{s}_{23}=\mathfrak{s}_{x},\mathfrak{s}_{31}=\mathfrak{s}_{y},\mathfrak{s}_{12}=\mathfrak{s}_{z}$. As usual, the summation is performed on the indices, the Latin indices take the values 1,2,3, the Greek ones - 0,1,2,3. The operator $\frac{\hbar}{2}\frac{i}{2}\mathfrak{s}_{kl}\sigma^{kl}$  is the operator of the projection of the spin of a  fermion at rest (or moving slowly) onto the direction $\bm{\mathfrak{s}}$.

For an observer moving with velocity $-\bm{v}$ relative to the fermion, the fermion has velocity $\bm{v}$ and is described by the equations that can be obtained from bispinor equations (1) - (2) by boost transformations:
\begin{gather}
e^{b_{0k}\sigma^{0k}}p_0\gamma^0e^{-b_{0k}\sigma^{0k}}e^{b_{0k}\sigma^{0k}}\psi^{(0)}=mce^{b_{0k}\sigma^{0k}}\psi^{(0)},\\
e^{b_{0k}\sigma^{0k}}\frac{\hbar}{2}\frac{i\mathfrak{s}_{kl}{2}}\sigma^{kl}e^{-b_{0k}\sigma^{0k}}e^{b_{0k}\sigma^{0k}}\psi^{(0)}=
\frac{\hbar}{2}e^{b_{0k}\sigma^{0k}}\psi^{(0)}.
\end{gather}
That is, by the equations
\begin{gather}
p_\alpha\gamma^\alpha\psi=mc\psi,\\
\frac{\hbar}{2}\frac{is_{\alpha\beta}\sigma^{\alpha\beta}}{2}\psi=\frac{\hbar}{2}\psi.
\end{gather}
We have
\begin{gather}
e^{b_{0k}\sigma^{0k}}=\cosh b+\frac{b_{0k}\sigma^{0k}}{b}\sinh b,\\
\frac{b_{01}}{b}=\frac{v_x}{v},\hspace{3mm}\frac{b_{02}}{b}=\frac{v_y}{v},\hspace{3mm}\frac{b_{03}}{b}=\frac{v_z}{v},
\hspace{3mm}\tanh 2b=\frac{v}{c},\\
\psi=e^{b_{0k}\sigma^{0k}}\psi^{(0)},\\
\frac{\hbar}{2}\frac{is_{\alpha\beta}\sigma^{\alpha\beta}}{2}
=\frac{\hbar}{2}e^{b_{0k}\sigma^{0k}}\frac{i\mathfrak{s}_{kl}\sigma^{kl}}{2}e^{-b_{0k}\sigma^{0k}}.
\end{gather}
The hypercomplex number $\mathfrak{s}_{\alpha\beta}\sigma^{\alpha\beta}$ when using the Lorentz transformations in the form (10) represents an antisymmetric tensor of the second order, similar to the electromagnetic field tensor $F_{\alpha\beta}$. More precisely, the tensor of the electromagnetic field, which in the fermion's own system is only a magnetic field, since $\mathfrak{s}_{0i}=0,\mathfrak{s}_{kl}\neq 0$. Let's call $\mathfrak{s}_{\alpha\beta}$ the spin tensor. The representation of the tensor $\mathfrak{s}_{\alpha\beta}$ in the form $\mathfrak{s}_{\alpha\beta}\sigma^{\alpha\beta}$ is similar to the representation of the 3-vector $(a_x,a_y,a_z)$ by quaternion $a_x\bm{i}+a_y\bm{j}+a_z\bm{k}$, or bivector $(a_x,a_y)$ with a complex number  $a_x+a_y i$. Note that, in addition to the spin tensor $\mathfrak{s}_{\alpha\beta}$, exists the spin pseudovector or the Pauli-Lubanski pseudovector \cite{l}
\begin{gather}
\lambda^\alpha=\varepsilon^{\alpha\beta\mu\nu}p_\beta s_{\mu\nu}.
\end{gather}
However, in this article we don't need it.

The constituents of the tensor $s_{\alpha\beta}$, defined by relation (10), can be conveniently written in 3-dimensional form as a 3-vector $\bm{s}_1$ and a 3-pseudovector $\bm{s}_2$, similarly to writing the tensor $F_{\alpha\beta}$ in the form of an electric field vector $\bm{E}$ and a pseudovector of a magnetic field $\bm{B}$:
\begin{gather}
\frac{\hbar}{2}i\frac{s_{\alpha\beta}\sigma^{\alpha\beta}}{2}=\frac{\hbar}{2}i(\bm{s}_1\bm{\varsigma}_1+\bm{s}_2\bm{\varsigma}_2),\\
\bm{s}_1\bm{\varsigma}_1=s_{01}\sigma^{01}+s_{02}\sigma^{02}+s_{03}\sigma^{03},\\
\bm{s}_2\bm{\varsigma}_2=s_{23}\sigma^{23}+s_{31}\sigma^{31}+s_{12}\sigma^{12},\\
\bm{s}_1=\frac{\bm{p}\times\bm{\mathfrak{s}}}{mc},\\
\bm{s}_2=\frac{p_0}{mc}\bm{\mathfrak{s}}-\frac{\bm{p(\bm{p\cdot\bm{\mathfrak{s}}})}}{mc(mc+p_0)}.
\end{gather}
Here we use 3-dimensional notation
\begin{gather}
\bm{\mathfrak{s}}=(\mathfrak{s}_x,\mathfrak{s}_y,\mathfrak{s}_z)=(\mathfrak{s}_{23},\mathfrak{s}_{31},\mathfrak{s}_{12}),\\
\bm{s}_1=(s_{1x},s_{1y},s_{1z})=(s_{01},s_{01},s_{03}),\\
\bm{s}_2=(s_{2x},s_{2y},s_{2z})=(s_{23},s_{31},s_{12}).
\end{gather}
Using (12) - (16) it is easy to be convinced that $p_\alpha\gamma^\alpha$ and $\frac{1}{2}s_{\mu\nu}\sigma^{\mu\nu}$ commute:
\begin{gather}
p_\alpha\gamma^\alpha\frac{1}{2}s_{\mu\nu}\sigma^{\mu\nu}-\frac{1}{2}s_{\mu\nu}\sigma^{\mu\nu}p_\alpha \gamma^\alpha=
p^\alpha s_{\alpha\beta}\gamma^\beta=0.
\end{gather}
According to equations (5) - (6), bispinor (9) describes a fermion with velocity $\bm{v}$, and operator (12) is the operator of the projection of the spin of a moving fermion. But in what direction is the spin projected in the system moving relative to the fermion? In this system, with the tensor $s_{\alpha\beta}$, two 3-directions  are connected: the direction of the vector $\bm{s}_1$ and the direction of the pseudovector $\bm{s}_2$. It is natural to assume that the spin is projected onto $\bm{s}_2$, since $\bm{s}_2$ tends to $\bm{\mathfrak{s}}_2$ as $\bm{p}\to 0$, while $\bm{s}_1$ tends to zero. The question is: what role does $\bm{s}_1$ play? Considering the spin group will help answer this question.

\section{The spin group}
Consider the transformation $e^{\frac{s_{\alpha\beta}\sigma^{\alpha\beta}}{2}\frac{\varphi}{2}}$, whose infinitesimal operator is $ \frac{is_{\alpha\beta}\sigma^{\alpha\beta}}{2}$. It is clear that this transformation does not change the spin projection operator (12):
\begin{gather}
e^{\frac{s_{\alpha\beta}\sigma^{\alpha\beta}}{2}\frac{\varphi}{2}}\,\frac{\hbar}{2}\frac{is_{\alpha\beta}\sigma^{\alpha\beta}}{2}\,  e^{-\frac{s_{\alpha\beta}\sigma^{\alpha\beta}}{2}\frac{\varphi}{2}}=\nonumber\\
\Bigl(\cosh s\frac{\varphi}{2}+\frac{\frac{s_{\alpha\beta}\sigma^{\alpha\beta}}{2}\frac{\varphi}{2}}{s\frac{\varphi}{2}}\sinh s\frac{\varphi}{2}\Bigr)
\frac{\hbar}{2}\frac{is_{\alpha\beta}\sigma^{\alpha\beta}}{2}\times\nonumber\\
\Bigl(\cosh s\frac{\varphi}{2}-\frac{\frac{s_{\alpha\beta}\sigma^{\alpha\beta}}{2}\frac{\varphi}{2}}{s\frac{\varphi}{2}}\sinh s\frac{\varphi}{2}\Bigr)=
\frac{\hbar}{2}\frac{is_{\alpha\beta}\sigma^{\alpha\beta}}{2}.
\end{gather}
Here $s=\sqrt{s_{\alpha\beta}\sigma^{\alpha\beta}s_{\alpha\beta}\sigma^{\alpha\beta}}=\sqrt{-\bm{\mathfrak{s}}^2}=i$. It is less clear that this transformation leaves unchanged not only operator (12), but also the 4-momentum of the fermion. It is not difficult to see that
\begin{gather}
e^{\frac{s_{\alpha\beta}\sigma^{\alpha\beta}}{2}\frac{\varphi}{2}}\,p_\mu\gamma^\mu\, e^{-\frac{s_{\alpha\beta}\sigma^{\alpha\beta}}{2}\frac{\varphi}{2}}=
(\cos\frac{\varphi}{2}+\frac{s_{\alpha\beta}\sigma^{\alpha\beta}}{2}\sin\frac{\varphi}{2})\times\nonumber\\
p_\mu\gamma^\mu(\cos\frac{\varphi}{2}-\frac{s_{\alpha\beta}\sigma^{\alpha\beta}}{2}\sin\frac{\varphi}{2})=p_\mu\gamma^\mu.
\end{gather}
Thus, this is a transformation belonging to the small Wigner group \cite{w} for the momentum $p_\mu$.

We apply the transformation $e^{\frac{s_{\alpha\beta}\sigma^{\alpha\beta}}{2}\frac{\varphi}{2}}$ to the eigen bispinor $\psi$ of the operator $ \frac{is_{\alpha\beta}\sigma^{\alpha\beta}}{2}$, i.e. to the solution $\psi$ of equation (6):
\begin{gather}
e^{\frac{s_{\alpha\beta}\sigma^{\alpha\beta}}{2}\frac{\varphi}{2}}\psi=\Bigl(\cosh s\frac{\varphi}{2}+
\frac{\frac{s_{\alpha\beta}\sigma^{\alpha\beta}}{2}\frac{\varphi}{2}}{s\frac{\varphi}{2}}\sinh s\frac{\varphi}{2}\Bigr)\psi=\nonumber\\
(\cos \frac{\varphi}{2}+\frac{s_{\alpha\beta}\sigma^{\alpha\beta}}{2}\sin \frac{\varphi}{2})\psi=
(\cos\frac{\varphi}{2}-i\sin\frac{\varphi}{2})\psi=\nonumber\\e^{-i\frac{\varphi}{2}}\psi.
\end{gather}
As we see, after such a transformation, the bispinor $\psi$ acquires the factor $e^{-i\frac{\varphi}{2}}$. This is due to the double-valuedness of the bispinor during rotations. After the rotation by $\varphi=2\pi$, the bispinor changes its sign, and transformation (23), as we will see below, contains a rotation. However, the bispinor $e^{-i\frac{\varphi}{2}}\psi$ describes a fermion with the same spin characteristics as the bispinor $\psi$. I.e., the transformation $e^{\frac{s_{\alpha\beta}\sigma^{\alpha\beta}}{2}\frac{\varphi}{2}}\psi$ leaves unchanged both the momentum of the fermion $\psi$ and its spin characteristics.
In this sense, it is the symmetry transformation of the fermion.

Note that when performing transformation  (22) (keeping the momentum unchanged) only relations  (12) - (16) are used, connecting $s_{\alpha\beta}$ with $\bm{\mathfrak{s}}$, and the condition $\bm{\mathfrak{s}}^2=1$. Therefore, all transformations  $e^{\frac{w_{\alpha\beta}\sigma^{\alpha\beta}}{2}\frac{\varphi}{2}}$ belong to the small Wigner group,
in which the parameters $s_{\alpha\beta}$  are replaced by the parameters $w_{\alpha\beta}$ connected with the {\em arbitrary} unit spatial 3-pseudovector $\bm{\mathfrak{w}}=(\mathfrak{w}_x,\mathfrak{w}_y,\mathfrak{w}_z)$  with the same relations as (15)-(16)
\begin{gather}
\bm{w}_1=\frac{\bm{p}\times\bm{\mathfrak{w}}}{mc},\\
\bm{w}_2=\frac{p_0}{mc}\bm{\mathfrak{w}}-\frac{\bm{p(\bm{p\cdot\bm{\mathfrak{w}}})}}{mc(mc+p_0)}.
\end{gather}
Here $\bm{w}_1=(w_{1x},w_{1y},w_{1z})= (w_{01},w_{02},w_{03})$,  $\bm{w}_2=(w_{2x},w_{2y},w_{2z})=(w_{23},w_{31},w_{12})$, $\bm{\mathfrak{w}}$ - is an arbitrary unit 3-pseudovector not related to spin in any way. However, these transformations, while conserving the momentum, do not preserve the spin properties of the fermion. Transformations  in which $w_{\alpha\beta}=s_{\alpha\beta}$ constitute a subgroup of the small Wigner group. We will call this subgroup the spin group. The transformations of the spin group, in contrast to other transformations of the small group, conserve not only the momentum, but also the spin characteristics of the fermion.

Transformations of the small group $e^{\frac{w_{\alpha\beta}\sigma^{\alpha\beta}}{2}\frac{\varphi}{2}}$ are neither spatial rotations nor boosts. However, they, like all the Lorentz transformations, can be represented as a sequence of two transformations - a spatial rotation and a boost
\begin{gather}
e^{\frac{w_{\alpha\beta}\sigma^{\alpha\beta}}{2}\frac{\varphi}{2}}=e^{b_{0k}\sigma^{0k}}e^{\frac{r_{kl}\sigma^{kl}}{2}}
\end{gather}
or boost and spatial rotation
\begin{gather}
e^{\frac{w_{\alpha\beta}\sigma^{\alpha\beta}}{2}\frac{\varphi}{2}}=e^{\frac{r'_{kl}\sigma^{kl}}{2}}e^{b'_{0k}\sigma^{0k}}.
\end{gather}
Performing this factorization, we obtain
\begin{gather}
e^{b_{ok}o^{ok}}=\nonumber\\
\sqrt{\frac{1+\bm{w}^2_2\tan^2\frac{\varphi}{2}}{1+\tan^2\frac{\varphi}{2}}}
\Bigl(1+\frac{\bm{w}_1\tan\frac{\varphi}{2}-(\bm{w}_1\times\bm{w}_2)\tan^2\frac{\varphi}{2}}{1+\bm{w}_2^2\tan^2\frac{\varphi}{2}}\bm{\varsigma}_1\Bigr)=\nonumber\\
\sqrt{\frac{1+\bm{w}^2_2\tan^2\frac{\varphi}{2}}{1+\tan^2\frac{\varphi}{2}}}
\Bigl(1+\frac{\bm{w}_1\tan\frac{\varphi}{2}+\frac{\bm{p}_\perp}{p_0}\bm{w}^2_2\tan^2\frac{\varphi}{2}}{1+\bm{w}_2^2\tan^2\frac{\varphi}{2}}\bm{\varsigma}_1\Bigr),
\end{gather}
\begin{gather}
e^{b'_{ok}o^{ok}}=\nonumber\\
\sqrt{\frac{1+\bm{w}^2_2\tan^2\frac{\varphi}{2}}{1+\tan^2\frac{\varphi}{2}}}
\Bigl(1+\frac{\bm{w}_1\tan\frac{\varphi}{2}+(\bm{w}_1\times\bm{w}_2)\tan^2\frac{\varphi}{2}}{1+\bm{w}_2^2\tan^2\frac{\varphi}{2}}\bm{\varsigma}_1\Bigr)=\nonumber\\
\sqrt{\frac{1+\bm{w}^2_2\tan^2\frac{\varphi}{2}}{1+\tan^2\frac{\varphi}{2}}}
\Bigl(1+\frac{\bm{w}_1\tan\frac{\varphi}{2}-\frac{\bm{p}_\perp}{p_0}\bm{w}^2_2\tan^2\frac{\varphi}{2}}{1+\bm{w}_2^2\tan^2\frac{\varphi}{2}}\bm{\varsigma}_1\Bigr),
\end{gather}
\begin{gather}
e^{\frac{1}{2}r_{kl}o^{kl}}=e^{\frac{1}{2}r'_{kl}o^{kl}}=\frac{1+\bm{w}_2\bm{\varsigma}_2\tan\frac{\varphi}{2}}{\sqrt{1+\bm{w}^2_2\tan^2\frac{\varphi}{2}}}.
\end{gather}
Here, to abbreviate the notation, we used the notation
\begin{gather}
\bm{a}\bm{\varsigma}_1=a_x\sigma^{01}+a_y\sigma^{02}+a_z\sigma^{03},\\
\bm{a}\bm{\varsigma}_2=a_x\sigma^{23}+a_y\sigma^{31}+a_z\sigma^{12},
\end{gather}
and $\bm{p}_\perp$ denotes the component of the vector $\bm{p}$ perpendicular to $\bm{w}_2$ and lying in the plane $(\bm{p},\bm{w}_2)$:
\begin{gather}
\bm{p}_\perp=\bm{p}-\frac{\bm{w}_2(\bm{w_2\cdot\bm{p}})}{\bm{w}^2_2}.
\end{gather}
It is easy to check that equalities (26) - (27) hold. Thus, the transformations of the small Wigner group are represented as a sequence of two transformations one with a unitary operator, the other with a Hermitian operator.

Unitary operators $e^{\frac{1}{2}r_{kl}o^{kl}}=e^{\frac{1}{2}r'_{kl}o^{kl}}$ carry out a spatial rotation around an arbitrary axis $\bm{w}_2$ by an angle
\begin{gather}
2r=2\arctan \Bigl(|\bm{w}_2|\tan\frac{\varphi}{2}\Bigr).
\end{gather}
In the partickle's own system, this is a rotation around an arbitrary axis $\bm{\mathfrak{w}}$ on an angle $\varphi$. This rotation changes the spin characteristics of the resting fermion, but does not change its 4-momentum $(mc,0,0,0)$. In the case of a moving particle, rotation around the $\bm{w}_2$ axis changes (rotates) the momentum component perpendicular to $\bm{w}_2$.

Hermitian operators $e^{b_{ok}o^{ok}}$ or $e^{b'_{ok}o^{ok}}$ perform boosts at the same speed $v=v'$
\begin{gather}
\tanh 2\frac{v}{c}=\tanh 2\frac{v'}{c}=
\Bigl|\frac{\bm{w}_1\tan\frac{\varphi}{2}+\frac{\bm{p}_\perp}{p_0}\bm{w}^2_2\tan^2\frac{\varphi}{2}}{1+\bm{w}_2^2\tan^2\frac{\varphi}{2}}\Bigr|=\nonumber\\=
\Bigl|\frac{\bm{w}_1\tan\frac{\varphi}{2}-\frac{\bm{p}_\perp}{p_0}\bm{w}^2_2\tan^2\frac{\varphi}{2}}{1+\bm{w}_2^2\tan^2\frac{\varphi}{2}}\Bigr|,
\end{gather}
but in different directions $\bm{w}_1\tan\frac{\varphi}{2}+\frac{\bm{p}_\perp}{p_0}\bm{w}^2_2\tan^2\frac{\varphi}{2}$ or $\bm{w}_1\tan\frac{\varphi}{2}-\frac{\bm{p}_\perp}{p_0}\bm{w}^2_2\tan^2\frac{\varphi}{2}$ perpendicular to $\bm{w}_2$.
In its own reference frame, $v=v'=0$ even these transformations  are absent, $e^{b_{ok}o^{ok}}=e^{b'_{ok}o^{ok}}=1$. In the case of a moving particle, these boosts change the component of the momentum perpendicular to $\bm{w}_2$ in such a way as to compensate for the change caused by the rotation.  As a result of two transformations - rotation and boost (or vice versa) - the momentum remains unchanged, but the spin characteristics change.

In order to return from the small Wigner group to the spin group, it is necessary in relations (26) - (36) to replace an arbitrary axis $\bm{\mathfrak{w}}_{\alpha\beta}$ by the axis $\bm{\mathfrak{s}}_{\alpha\beta}$ onto which the spin of the fixed fermion is projected.

In the case of a fermion at rest, we see that the transformation of the spin group is a spatial rotation of the system through an arbitrary angle $\varphi$ around the $\bm{\mathfrak{s}}$ axis, onto which the spin of the resting fermion is projected. This transformation does not change the spin characteristics and does not change the 4-momentum of the resting fermion $(mc,0,0,0)$. When the parameter $\varphi$ changes continuously, the turn becomes rotation. The representation of a fermion as a rotating particle reflects this possibility of rotation of the fermion by an arbitrary angle around the $\bm{\mathfrak{s}}$ axis without changing the properties of the fermion.

In the case of a moving fermion, the transformation of the spin group, which does not change the properties of the fermion, is a sequence of spatial rotation and a boost coordinated with this rotation (or vice versa).

Spatial rotation is performed at an angle
\begin{gather}
2r=2\arctan \Bigl(|\bm{s}_2|\tan\frac{\varphi}{2}\Bigr)
\end{gather}
around the $\bm{s}_2$ axis, which is the $\bm{\mathfrak{s}}$ axis transformed into a moving system. If the parameter $\varphi$ changes continuously, instead of a turn, we obtain rotation around the axis onto which the spin of the moving fermion is projected.

This turn, in the case of a moving fermion, is accompanied by boosts in the direction $\bm{s}_1\tan\frac{\varphi}{2}+\frac{\bm{p}_\perp}{p_0}(1+\bm{s}^2_1)\tan^2\frac{\varphi}{2}$ or $\bm{s}_1\tan\frac{\varphi}{2}-\frac{\bm{p}_\perp}{p_0}(1+\bm{s}^2_1)\tan^2\frac{\varphi}{2}$, perpendicular to $\bm{s}_2$. The components of these directions are determined by the vector $\bm{s}_1$ perpendicular to $\bm{s}_2$ and the component of the momentum perpendicular to $\bm{p}_\perp$.
The speed of the boosts is the same, $v=v'$. It is determined from the equation
\begin{gather}
\tanh 2\frac{v}{c}=\tanh 2\frac{v'}{c}=
\Bigl|\frac{\bm{s}_1\tan\frac{\varphi}{2}+\frac{\bm{p}_\perp}{p_0}(1+\bm{s}^2_1)\tan^2\frac{\varphi}{2}}{1+(1+\bm{s}_1^2)\tan^2\frac{\varphi}{2}}\Bigr|=
\nonumber\\\Bigl|\frac{\bm{s}_1\tan\frac{\varphi}{2}-\frac{\bm{p}_\perp}{p_0}(1+\bm{s}^2_1)\tan^2\frac{\varphi}{2}}{1+(1+\bm{s}_1^2)\tan^2\frac{\varphi}{2}}\Bigr|.
\end{gather}
Since the boosts are changes in velocity, for a moving fermion the rotations around the $\bm{s}_2$ axis are complemented by changes in velocity perpendicular to this axis. The velocities $\bm{v}$ or $\bm{v}'$ are periodic functions of the parameter $\varphi$. Therefore, continuous changes in the parameter $\varphi$, which do not change the properties of the fermion, correspond to rotation around the $\bm{s}_2$ axis and the oscillating motions associated with them, perpendicular to this axis $\bm{s}_2$.

As we see, in the depiction reflecting the symmetry of a moving fermion, it is not only a rotating fermion, but also an oscillating one perpendicular to the rotation axis.
Of course, the depiction of a rotating particle is only an image that helps to relate the existence of spin with the classical ideas of angular momentum. The concept of an oscillating motion perpendicular to the axis of rotation is a similar image. The question arises: the existence of which physical characteristic reflects the image of the oscillating motion?

Let us turn to experiments that directly convince us of the existence of spin in a charged particle. These are experiments similar to that of Stern-Gerlach experiment \cite{gs1}, \cite{gs2}. That is, observing the interaction of the spin-related intrinsic magnetic moment of a particle with an external magnetic field. Since the directions of the spin and magnetic moment are related to the direction of the component $\bm{s}_2$ of the spin tensor $s_{\alpha\beta}$, it is this component that is observed in such experiments.

In a similar way, the observation of the polarization of a moving charged particle with spin in an electric field convinces us of the existence of an electric moment of such a particle. This kind of polarization accompanies spin-orbit interaction or electron scattering at the Coulomb center \cite{bd}. Usually, to explain it, the process is considered in the particle's own system. In this system, the particle is stationary, and the external electric field generates an induction magnetic field, which interacts with the magnetic moment of the particle at rest and polarizes it. If we remain in the nuclear reference frame or Coulomb center system, in which there is only an electric field, then the polarization can be explained only by the appearance of an electric dipole moment in the moving particle. Normally, it is generated by the magnetic moment of a stationary particle in the transition from its own system to the system in which it moves. It is perpendicular to the magnetic moment of the particle. It is similar to the generation of an induction electric field  in a system moving  relative to the magnetic field. Since the spin tensor $s_{\alpha\beta}$ and the electromagnetic field tensor $F_{\alpha\beta}$ are transformed in the same way, these two ways of explaining the polarization of a moving fermion in an electric field are consistent and equivalent.

Returning to the oscillating motion of a moving charged particle with spin, we can consider it as an image that helps us understand the appearance of an electric moment in such a particle, just as the image of a rotating charged particle does so with respect to the magnetic moment.

\section{Acknowledgements}
The authors would like to thank  Prof. Denis Kovalenko for stimulating discussions.

\section{Summary}
In this way we have shown that there is a symmetry group of moving fermions with spin $\frac{\hbar}{2}$. This is a spin group. The transformations of this group do not change the fermion momentum and its spin characteristics. These transformations are sequences of a rotation and a boost coordinated with it (or vice versa). This spin group is a subgroup of the small Wigner group. The spin projection operator $\frac{\hbar}{2}\frac{is_{\alpha\beta}\sigma^{\alpha\beta}}{2}$ is the infinitesimal operator of this spin group, that is, the symmetry group of the moving fermion.

\end{document}